\documentclass[a4paper,11pt]{article}
\pdfoutput=1 % if your are submitting a pdflatex (i.e. if you have
\usepackage{jheppub} % for details on the use of the package, please
\usepackage{braket}
\usepackage{float}

\title{\boldmath More on the Bending of Light in Quantum Gravity}

\author{Dong Bai}
\author{and}
\author{Yue Huang}

\affiliation[1]{Key Laboratory of Theoretical Physics, Institute of Theoretical Physics,\\Chinese Academy of Sciences, Beijing 100190, China}
\affiliation[2]{School of Physical Sciences, University of Chinese Academy of Sciences,\\No.19A Yuquan Road, Beijing 100049, China}

\emailAdd{dbai@itp.ac.cn}
\emailAdd{huangyue@itp.ac.cn}

\abstract{We reconsider the long-range effects of the scattering of massless scalars and photons from a massive scalar object in quantum gravity. At the one-loop level, the relevant quantum mechanical corrections could be sorted into the graviton double-cut contributions, massless-scalar double-cut contributions and photon double-cut contributions. In Ref.~\cite{Bjerrum-Bohr:2014zsa,Bjerrum-Bohr:2016hpa} N.E.J.~Bjerrum-Bohr et al.~have considered explicitly the implications of the graviton double-cut contributions on the gravitational bending of light and some classical formulations of the equivalence principle, using the modern double-copy constructions and on-shell unitarity techniques. In this article, instead we consider all three contributions and redo the analysis using the traditional Feynman diagrammatic approach. Our results on the graviton double-cut contributions agree with the aforementioned references, which acts as a nontrivial check of previous computations. Furthermore, it turns out that the massless-scalar double-cut contributions and the photon double-cut contributions do leave non-vanishing quantum effects on the scattering amplitudes and the gravitational bending of light. Yet, we find that the general structure of the gravitational amplitudes and the quantum discrepancy of the equivalence principle suggested in the aforementioned references remain intact.}

\begin{document} 
\maketitle
\flushbottom

\section{Introduction} 

The reconciliation of quantum mechanics and general relativity has been a long-standing open question in theoretical physics since the beginning of last century, and different proposals have been put forward. For instance, recently a possible unification framework based on spin and scaling gauge invariance has been proposed in Ref.~\cite{Wu:2015wwa,Wu:2015hoa}. At present, there is no general consensus on the ultimate solution yet. But still, the long-range effects of the underlying quantum gravity could now be calculated reliably by treating general relativity as a low-energy effective field theory \cite{Donoghue:1993eb,Donoghue:1994dn} (see also Ref.~\cite{Muzinich:1995uj,Hamber:1995cq,Akhundov:1996jd,Kazakov:2000mu,BjerrumBohr:2002sx,BjerrumBohr:2002kt,BjerrumBohr:2002ks,Khriplovich:2002bt,Khriplovich:2004cx,Satz:2004hf,Holstein:2008sx,Park:2010pj,Marunovic:2011zw,Marunovic:2012pr,Burns:2014bva,Frob:2016xte} for later developments), even though general relativity is non-renormalizable and requires an infinite number of counterterms to absorb all the ultraviolet divergences. For a recent review, we recommend Ref.~\cite{Donoghue:2017pgk}. It is shown in Ref.~\cite{BjerrumBohr:2002kt,Holstein:2008sx,Bjerrum-Bohr:2013bxa} that the spin-independent part of the quantum Newtonian potential between a small mass and a large mass looks like
\begin{equation}
V(r)=-\frac{GMm}{r}\left(1+\frac{3G(M+m)}{r}+\frac{41G\hbar}{10r^2}\right),
\label{QMCNMassive}
\end{equation}
where $M$ is a large scalar object, say, the Sun, $m$ is the small test mass whose spin could be $0$, $1/2$ or $1$, and $r$ is the relative distance between these two objects. $G$ and $\hbar$ denote Newton's constant and the Planck constant, respectively. 

It is of physical interest to study other long-range effects of quantum gravity, among which the leading quantum corrections to the gravitational bending of light\footnote{In this article, by ``the gravitational bending of light'', what we really mean is the gravitational bending of massless particles, including not only the photon, but also the hypothetical massless scalar particle, etc.} around the Sun is a perfect target for attack. In order to handle this problem in the framework of quantum field theory, the Sun is mimicked by a heavy scalar field. As shown later, this approximation gives the correct classical bending angles. With this in mind, one could consider the Einstein-Hilbert action coupled to photons and two neutral scalars, one massless and the other massive
\begin{equation}
S=\int\mathrm{d}^4x\sqrt{-g}\left[-\frac{2}{\kappa^2}R-\frac{1}{4}\left(\nabla_{\mu} A_{\nu}-\nabla_{\nu}A_{\mu}\right)^2+\frac{1}{2}(\partial_\mu\varphi)^2+\frac{1}{2}(\partial_{\mu}\Phi)^2-\frac{1}{2}M^2\Phi^2\right].
\label{Action}
\end{equation}
In this article, we adopt the mostly minus metric signature $(+,-,-,-)$ and $\kappa^2=32\pi G$. $\nabla_{\mu}$ is the usual covariant derivative with $\nabla_{\mu}A^{\nu}=\partial_{\mu}A^{\nu}+\Gamma^{\nu}_{\ \mu\lambda}A^{\lambda}$. These fields are denoted in the rest part of this article as follows: graviton $h$, photon $\gamma$, massless scalar $\varphi$, massive scalar $\Phi$, graviton Faddeev-Popov (FP) ghost $c$ (not shown explicit in the above action). Although there might not be a massless scalar particle in the real nature, the massless scalar $\varphi$ is introduced for the sake of studying the impact of quantum corrections on the some classical formulations of the equivalence principle (see Section \ref{BOL}).

To calculate the leading quantum corrections to the bending angles of the massless scalar $\varphi$ and photon $\gamma$ in the gravitational field of the massive field $\Phi$, one needs to first calculate the relevant scattering amplitudes up to one loop. As microscopically it is the scattering process $\varphi(\gamma)\Phi\to\varphi(\gamma)\Phi$ that is under consideration, there are only $t$-channel contributions. The $s$-channel and $u$-channel Feynman diagrams simply do not exist. Generally, for the scattering of a test particle in the gravitational field of a large mass, a typical scattering amplitude up to one loop looks like
\begin{equation}
\mathcal{M}\sim A+B q^2+\cdots+\alpha\kappa^4\frac{1}{q^2}+\beta_1\kappa^4\ln(-q^2)+\beta_2\kappa^4\frac{1}{\sqrt{-q^2}} + \gamma \kappa^4 \ln \left(1+\frac{q^2}{m^2}\right)+\cdots,
\end{equation}
where $q$ is the momentum transfer. Among these terms, only the \emph{gapless} non-analytic contributions, whose branch-cuts extend to the origin of the complex plane, would lead to long-range effects after Fourier transformations. The analytic and gapped non-analytic corrections will yield short-range $\delta^3(r)$ and $\exp (-mr)$ effects in the potential. Therefore, in order to work out the long-range effects of the one-loop quantum corrections, one only needs to consider diagrams with $t$-channel massless double-cuts \cite{BjerrumBohr:2002sx,BjerrumBohr:2002kt,Bjerrum-Bohr:2014zsa,Bjerrum-Bohr:2013bxa,Bjerrum-Bohr:2016hpa}.

For scattering processes between $\varphi(\gamma)$ and $\Phi$, the relevant Feynman diagrams with $t$-channel massless double-cuts could be sorted into three classes: the graviton double-cut contributions, the massless-scalar double-cut contributions and the photon double-cut contributions, whose meanings will be clear later on. In Ref.~\cite{Bjerrum-Bohr:2014zsa,Bjerrum-Bohr:2016hpa}, the graviton double-cut contributions have been calculated using the modern on-shell unitarity techniques and the double-copy constructions. In this article, we shall consider the implications of all three contributions. Compared to Ref.~\cite{Bjerrum-Bohr:2014zsa,Bjerrum-Bohr:2016hpa}, instead we adopt the traditional Feynman diagrammatic approach. As shown in later sections, our Feynman diagrammatic calculations of the graviton double-cut diagrams match the results of Ref.~\cite{Bjerrum-Bohr:2014zsa,Bjerrum-Bohr:2016hpa} exactly, which could be viewed as a nontrivial check of previous calculations. On the other hand, our results on the net effects of all three contributions  are new, and could be useful for further studies in this direction.   

The following parts of this article are organized as follows: In Section \ref{TLA}, we present the tree-level amplitudes for massless scalar and photon, and provide a concise description of how to extract the Newtonian potential from scattering amplitudes. In Section \ref{OLA}, we calculate the relevant one-loop amplitudes using Feynman diagrams. The main results are presented in terms of a set of parameters inspired by the modern one-loop calculation techniques. Then, we derive the low-energy limit of the obtained one-loop amplitudes. In Section \ref{BOL}, we calculate the bending of light in quantum gravity, and comment on the validity of quantum violation of some classical formulations of the equivalence principle. In Section \ref{FD}, we make brief remarks on possible further directions. Especially, we find it physically interesting to calculate quantum corrections to other classical observables of general relativity, such as the Shapiro time delay effect, the so-called fourth test of general relativity. We also attach two appendices at the end of this article to provide extra details of our computations. The Mathematica codes for the one-loop quantum gravity calculations carried out in this article could be found in the supplement materials.

\section{Tree-Level Amplitudes}
\label{TLA}
In this section, we extract the Newtonian potential from the tree-level scattering amplitudes. The model under investigation is given by Eq.~\eqref{Action}
\begin{equation}
S=\int\mathrm{d}^4x\sqrt{-g}\left[-\frac{2}{\kappa^2}R-\frac{1}{4}\left(\nabla_{\mu} A_{\nu}-\nabla_{\nu}A_{\mu}\right)^2+\frac{1}{2}(\partial_\mu\varphi)^2+\frac{1}{2}(\partial_{\mu}\Phi)^2-\frac{1}{2}M^2\Phi^2\right],\tag{\ref{Action}}
%\label{Action}
\end{equation}
where $A_{\mu}$ is the photon field, $\varphi$ and $\Phi$ are the massless and massive scalar field respectively. To do the perturbative calculations, one could write the metric as $\eta_{\mu\nu}+\kappa h_{\mu\nu}$, and expand all terms in $h_{\mu\nu}$. After quantization, $h_{\mu\nu}$ gives rise to the massless spin-$2$ graviton. 

With the above action, the tree-level scattering amplitudes of the massless scalar $\varphi$ and the photon $\gamma$ in the gravitational field of the massive scalar $\Phi$ could be worked out straightforwardly, and are presented as follows:
\begin{itemize}
\item{Massless scalar $\varphi$:}
\begin{align}
\mathcal{M}^{(0)}(\phi(p_1)\Phi(p_2)\phi(p_3)\Phi(p_4))=\frac{\kappa^2}{4}\frac{(s-M^2)(u-M^2)}{t}.
\end{align}

\item{Photon $\gamma$:}
\begin{align}
&\mathcal{M}^{(0)}(\gamma(p_1^+)\Phi(p_2)\gamma(p_3^+)\Phi(p_4))=0, \quad \mathcal{M}^{(0)}(\gamma(p_1^-)\Phi(p_2)\gamma(p_3^-)\Phi(p_4))=0,\\
&\mathcal{M}^{(0)}(\gamma(p_1^+)\Phi(p_2)\gamma(p_3^-)\Phi(p_4))=\frac{\kappa^2}{4}\frac{\bra{p_3|p_2}p_1]^2}{t},\\
&\mathcal{M}^{(0)}(\gamma(p_1^-)\Phi(p_2)\gamma(p_3^+)\Phi(p_4))=\frac{\kappa^2}{4}\frac{\bra{p_1|p_2}p_3]^2}{t}.
\label{PhotonMP}
\end{align}
\end{itemize}
In the above, we have adopted the in-in formalism with all momenta defined as incoming, as well as the spinor-helicity variables. The Mandelstam variables $s$, $t$ and $u$ are given by $s=(p_1+p_2)^2$, $t=(p_1+p_3)^2$ and $u=(p_1+p_4)^2$, respectively, and satisfy the identity $s+t+u=2M^2$. 

There are various definitions of quantum gravity potential in the literature, depending on the physical situations to be handled, the Feynman diagrams involved, etc. A physical plausible definition should certainly be gauge independent \cite{Kazakov:2000mu}. In this article, we shall take the definition of the (quantum) Newtonian potential given by Eq.~\eqref{DefinitionNewtonianPotential1}, Eq.~\eqref{DefinitionNewtonianPotential2} and Eq.~\eqref{DefinitionNewtonianPotential3}, which are used by various authors \cite{Hamber:1995cq,BjerrumBohr:2002sx,BjerrumBohr:2002kt,BjerrumBohr:2002ks,Bjerrum-Bohr:2014zsa,Bjerrum-Bohr:2016hpa}. Such construction relates the potential directly to the full scattering amplitudes and therefore enjoys an explicit gauge independence. In the low-energy limit, $t\to -\textbf{q}^2$, $s\to M^2+2M\omega$, we have
\begin{itemize}
\item{Massless scalar $\varphi$:}
\begin{align}
\mathcal{M}^{(0)}(\phi(p_1)\Phi(p_2)\phi(p_3)\Phi(p_4))=\frac{32\pi GM^2\omega^2}{\textbf{q}^2} \Longrightarrow \tilde{V}^{(0)}_{\varphi}(\textbf{q})=-\frac{8\pi GM\omega}{\textbf{q}^2}.
\end{align}
When Fourier transformed back to the coordinate space, we get the Newtonian potential
\begin{align}
V^{(0)}_{\varphi}(r)=\int\frac{\mathrm{d}^3\textbf{q}}{(2\pi)^3}\exp(i\textbf{q}\cdot\textbf{r})\tilde{V}^{(0)}_{\varphi}(\textbf{q})=-\frac{2GM\omega}{r}.
\end{align}
\item{Photon $\gamma$:}
The tree-level amplitudes for the photon shown above, especially the spinor chains, are displayed in terms of the in-in formalism. When translated into the standard in-out formalism (with $p_3\to-p_3$), one has
\begin{align}
&\mathcal{M}^{(0)}(\gamma(p_1^+)\Phi(p_2)\gamma(p_3^-)\Phi(p_4))=-\frac{\kappa^2}{4}\frac{\bra{p_3|p_2}p_1]^2}{t},\\
&\mathcal{M}^{(0)}(\gamma(p_1^-)\Phi(p_2)\gamma(p_3^+)\Phi(p_4))=-\frac{\kappa^2}{4}\frac{\bra{p_1|p_2}p_3]^2}{t},
\label{PhotonMP}
\end{align}
and the Mandelstam variable $t=(p_1-p_3)^2$. In the low-energy limit, besides $t\to -\textbf{q}^2$, $s\to M^2+2M\omega$, one has $\bra{p_3|p_2}p_1]^2 (\bra{p_1|p_2}p_3]^2)\to 4M^2\omega^2$. As a result, the low-energy limits of the relevant scattering amplitudes are given by
\begin{align}
&\mathcal{M}^{(0)}(\gamma(p_1^+)\Phi(p_2)\gamma(p_3^-)\Phi(p_4))=\frac{32\pi GM^2\omega^2}{\textbf{q}^2},\\
&\mathcal{M}^{(0)}(\gamma(p_1^-)\Phi(p_2)\gamma(p_3^+)\Phi(p_4))=\frac{32\pi GM^2\omega^2}{\textbf{q}^2},\\
&\Longrightarrow  \tilde{V}^{(0)}_{\gamma}(\textbf{q})=-\frac{8\pi GM\omega}{\textbf{q}^2}.
%\label{PhotonMP}
\end{align}
The Newtonian potential in the coordinate space is then given by $V^{(0)}_{\gamma}(r)=-2GM\omega/r$, which is symbolically the same as that of the massless scalar $\varphi$. This is nothing but a direct manifestation of the classical equivalence principle.
\end{itemize} 

\section{One-Loop Amplitudes}
\label{OLA}
\begin{figure}
\centering
\includegraphics[scale=1]{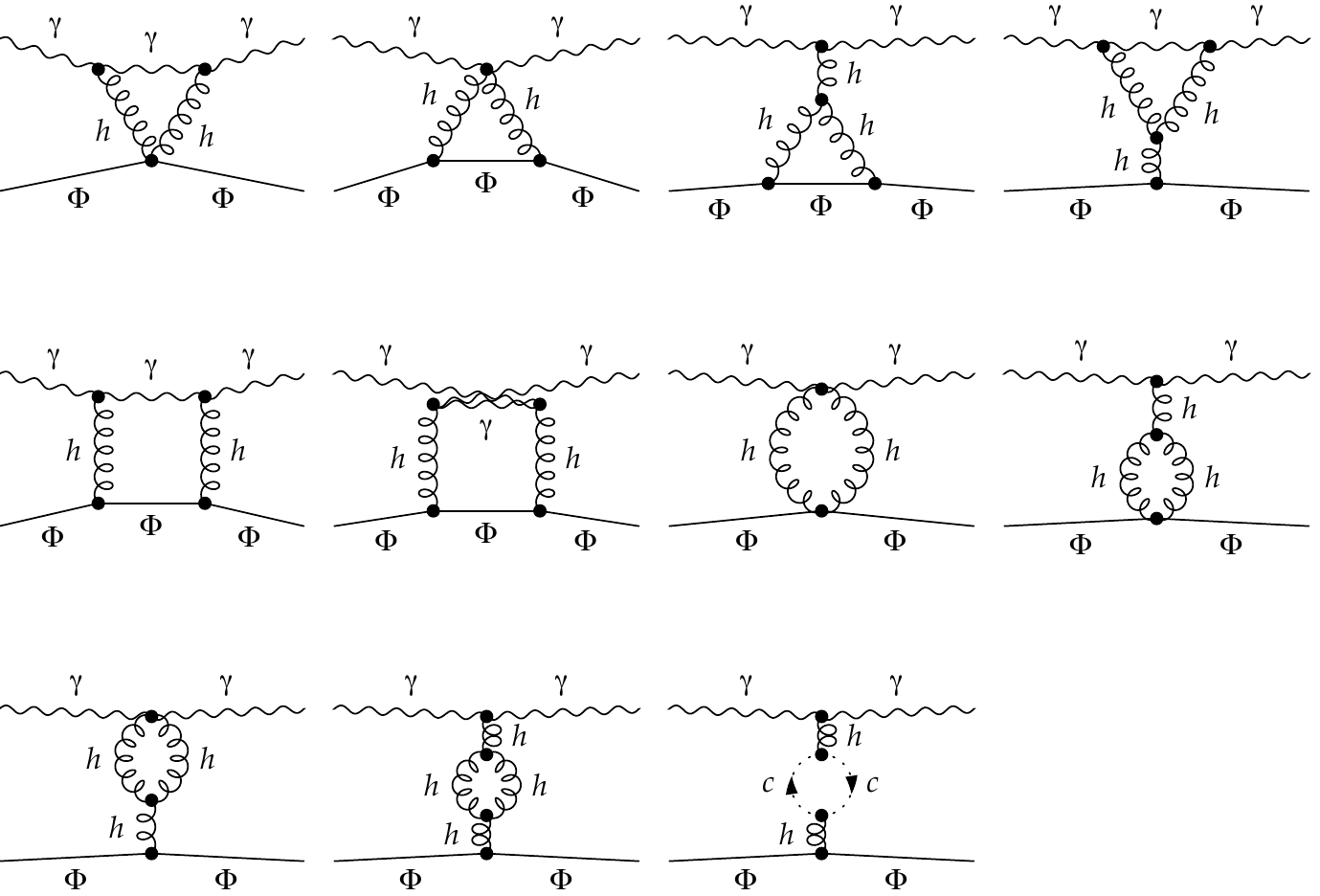}
\caption{Graviton double-cut diagrams for the one-loop scattering between the photon $\gamma$ and the massive scalar $\Phi$. Unlike Ref.~\cite{Bjerrum-Bohr:2014zsa,Bjerrum-Bohr:2016hpa}, in this article we take the time direction to be horizontal. In the last diagram, we have the graviton FP ghost running in the loop.}
\label{pic1}
\end{figure}

\begin{figure}
\centering
\includegraphics[scale=1]{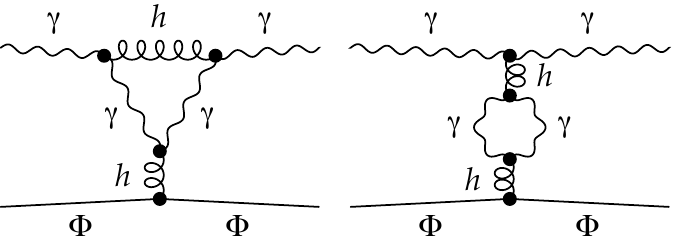}
\caption{Photon double-cut diagrams for the one-loop scattering between the photon $\gamma$ and the massive scalar $\Phi$.}
\label{pic2}
\end{figure}

\begin{figure}
\centering
\includegraphics[scale=1]{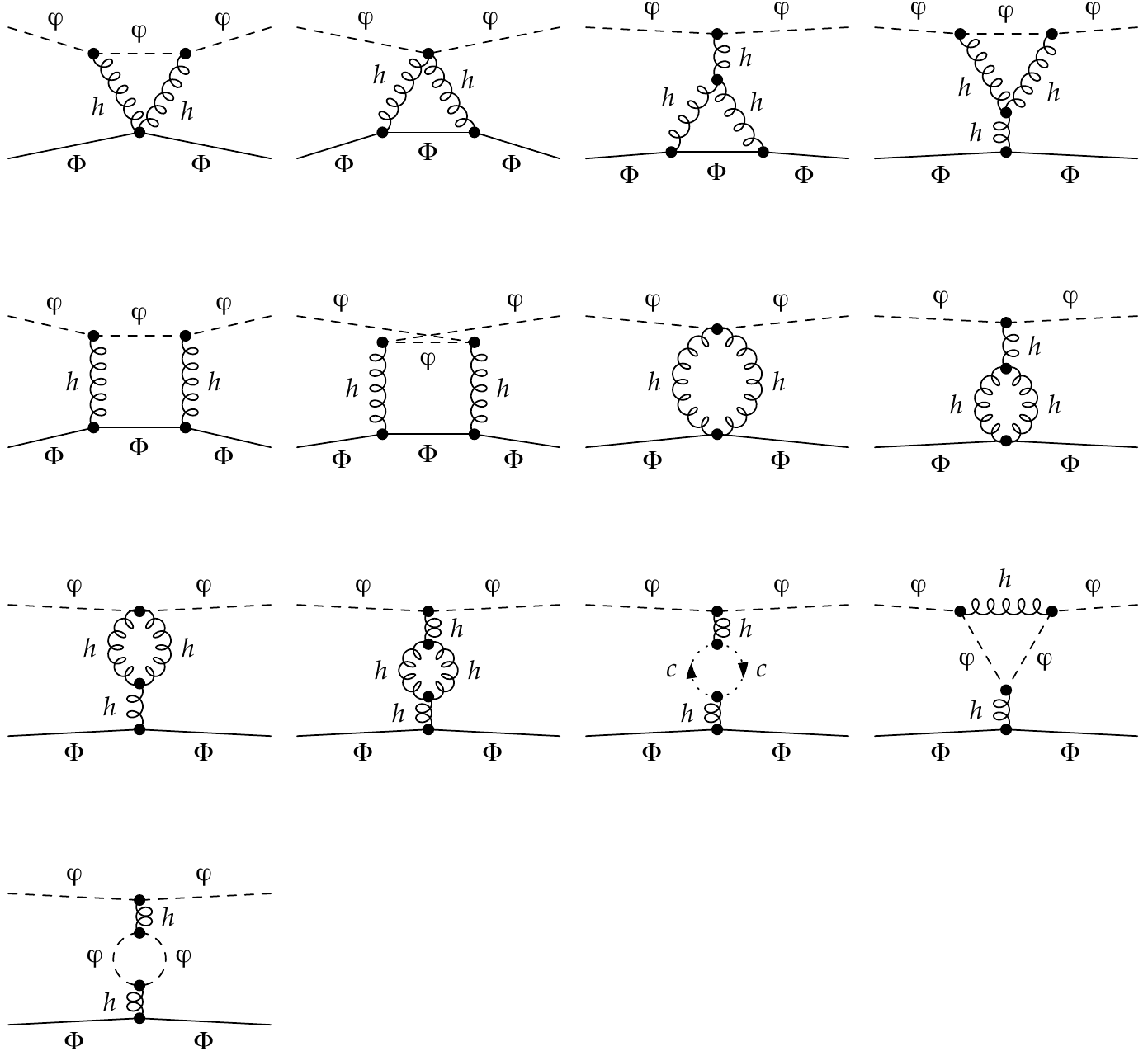}
\caption{The one-loop Feynman diagrams for the scattering of the massless scalar $\varphi$ from the massive scalar $\Phi$. The first eleven diagrams constitute the graviton double-cut contributions, while the last two diagrams constitute the massless-scalar double-cut contributions.}
\label{pic3}
\end{figure}

In this section, we move on to calculate the relevant one-loop amplitudes. As mentioned in Introduction, the one-loop scatterings of $\varphi$ and $\gamma$ in the gravitational field of the massive object $\Phi$ could be sorted into three classes: the graviton double-cut diagrams, the massless-scalar double-cut diagrams and the photon double-cut diagrams. Let's take the photon as an example to show our classifications. The relevant Feynman diagrams could be found in Fig.~\ref{pic1} and Fig.~\ref{pic2}. Fig.~\ref{pic1} is characterized by the feature that one could find a horizontal cut for every diagram such that two graviton propagators are cut down. Here we have included also the graviton FP ghost $c$ in the graviton double-cut diagrams. As a result, we shall call diagrams in Fig.~\ref{pic1} as the graviton double-cut diagrams. Similarly, diagrams in Fig.~\ref{pic2} are named as the photon double-cut diagrams, as for these diagrams one could find a horizontal cut such that two photon propagators are cut down. Here we have omitted the massless-scalar double-cut diagrams for the photon scattering, i.e., the Feynman diagram with the massless scalar $\varphi$ running in the vacuum polarization loop. This is because unlike the photon the massless scalar $\varphi$ is an ``optional" ingredient for our real world. Similarly, for the massless-scalar scattering, we will omit the Feynman diagram with the photon $\gamma$ in the vacuum polarization loop and consider only the graviton double-cut diagrams and the massless-scalar double-cut diagrams (Fig.~\ref{pic3}).\footnote{We would like to thank Prof.~N.E.J.~Bjerrum-Bohr, Prof.~John F.~Donoghue, Prof.~Barry R.~Holstein, Dr.~Ludovic Plant\'e, and Prof.~Pierre Vanhove for this suggestion.}

Unlike Ref.~\cite{Bjerrum-Bohr:2014zsa,Bjerrum-Bohr:2016hpa} the one-loop calculations are done using the traditional Feynman diagrammatic approach rather than the modern on-shell unitarity techniques. There are several reasons for this choice. First, this allows us to carry out independent calculations of bending of light in quantum gravity, and could be viewed as a useful complement to the existing literature. Second, there are various well-developed public Mathematica codes, such as FeynRules \cite{Alloul:2013bka}, FeynArts \cite{Hahn:2000kx}, FormCalc \cite{Hahn:2016ebn}, etc, which semi-automatize the one-loop Feynman diagrammatic calculations. 

Concerning the representation of our results, we utilize the fact that a generic one-loop integral could be represented as $\mathcal{M}=\sum_i c_i I_i$, where $c_i$ are rational functions of various kinematic invariants and $I_i$ are some known scalar integral functions representing simple one-loop diagrammatic contributions such as box, triangle and bubble diagrams. This result goes back to Passarino and Veltman \cite{Passarino:1978jh}. As a result, for the problem of the gravitational bending of light, the gapless non-analytic parts of the one-loop amplitudes, the only parts that contribute to the long-range quantum gravity effects, could be parametrized in the following way
\begin{align}
&\mathcal{M}^{(1)}_{\eta}(p_1,p_2,p_3,p_4)|_{\text{non-analytic}}\sim\text{Bo}^{\eta}\times I_4(t,s)+\text{Bo}'^{\eta}\times I_4(t,u)+\text{T}^{\eta}\times I_3(t,0)\nonumber\\ 
&+ \text{T}'^{\eta}\times I_3(t,M^2)+\text{Bu}^{\eta}\times I_2(t,0).
%\label{OneLoopTensorReduction}
\end{align}
$I_4(t,s) (I_4(t,u))$, $I_3(t,0)$, $I_3(t,M^2)$ and $I_2(t,0)$ are the standard scalar integrals, whose explicit expressions are found in Appendix \ref{UI}. The $\text{Bo}^{\eta}$, $\text{T}^{\eta}$, $\text{T}'^{\eta}$ and $\text{Bu}^{\eta}$ parameters are meromorphic functions of kinematic invariants, such as the Mandelstam variables $s$, $t$ and $u$.

For the massless scalar $\varphi$, the one-loop coefficients are given by
\begin{itemize}
\item{The box coefficients:}
\begin{align}
\text{Bo}^{\varphi}=\frac{\kappa^4}{256\pi^2}(M^2-s)^4,
\end{align}
\begin{align}
\text{Bo}'^{\varphi}=\frac{\kappa^4}{256\pi^2}(-M^2+s+t)^4,
\end{align}

\item{The triangle coefficients:}
\begin{align}
\text{T}^{\varphi}=\frac{\kappa^4}{256 \pi^2} t (-2 M^2 + 2 s + t)^2,
\end{align}
\begin{align}
&\text{T}'^{\varphi}=\frac{\kappa^4}{256 \pi^2 (-4 M^2 + t)^2}\Big[-60 M^{10} + 2 M^8 (60 s + 73 t) + t^3 (3 s^2 + 3 s t + t^2)\nonumber\\
&- 20 M^6 (3 s^2 + 12 s t + 7 t^2) - M^2 t^2 (30 s^2 + 36 s t + 13 t^2)\nonumber\\
&+ 3 M^4 t (30 s^2 + 50 s t + 21 t^2)\Big],
\end{align}

\item{The bubble coefficient:}
\begin{align}
&\text{Bu}^{\varphi}=\frac{\kappa^4}{15360 \pi^2 (-4 M^2 + t)^2} \Big[2968 M^8 - 424 M^6 (14 s + 9 t)\nonumber\\ 
&+ t^2 (103 s^2 + 103 s t + 23 t^2)-M^2 t (1064 s^2 + 1270 s t + 341 t^2) \nonumber\\ 
&+M^4 (2968 s^2 + 5096 s t + 1787 t^2)\Big],
\end{align}

\end{itemize}

The one-loop coefficients for the photon $\gamma$ with the $(-+)$ helicity are given by
\begin{itemize}
\item{The box coefficients:}
\begin{align}
&\text{Bo}^{\gamma}=\frac{\kappa^4 \bra{p_1}p_2|p_3]^2 }{512 \pi^2 (M^4 - 2 M^2 s + s (s + t))^2}(M^2 - s)^2 \Big[2 M^8 - 8 M^6 s - 4 M^2 s^2 (2 s + t)\nonumber\\
& + 2 M^4 s (6 s + t) + s^2 (2 s^2 + 2 s t + t^2)\Big],
\end{align}
\begin{align}
&\text{Bo}'^{\gamma}=\frac{\kappa^4 \bra{p_1}p_2|p_3]^2}{512 \pi^2 (M^4 - 2 M^2 s + s (s + t))^2}(-M^2 + s + t)^2 \Big[t^2 (-2 M^2 + s + t)^2 \nonumber\\
&+ 2 (-M^2 + s + t)^2 (M^4 - 2 M^2 s + s (s + t))\Big],
\end{align}

\item{The triangle coefficients:}
\begin{align}
&\text{T}^{\gamma}=\frac{\kappa^4 \bra{p_1}p_2|p_3]^2}{512 \pi^2 (M^4 - 2 M^2 s + s (s + t))^2} t \Big[8 M^8 + 8 s^4 + 16 s^3 t + 13 s^2 t^2\nonumber\\
& + 5 s t^3 + t^4 - 2 M^6 (16 s + 5 t) + 2 M^4 (24 s^2 + 18 s t + 5 t^2)\nonumber\\
& - M^2 (32 s^3 + 42 s^2 t + 20 s t^2 + 5 t^3)\Big],
\end{align}
\begin{align}
&\text{T}'^{\gamma}=-\frac{\kappa^4\bra{p_1}p_2|p_3]^2 }{512 \pi^2 (-4 M^2 + t)^2 (M^4 - 2 M^2 s + s (s + t))^2}\Big[120 M^{14} - 20 M^{12} (24 s  \nonumber\\
&+ 17 t)+ 4 M^{10} (180 s^2 + 320 s t + 131 t^2) - 8 M^8 (60 s^3 + 215 s^2 t + 195 s t^2 + 62 t^3)\nonumber\\
& - t^3 (6 s^4 + 12 s^3 t + 11 s^2 t^2 + 5 s t^3 + t^4) + M^2 t^2 (60 s^4 + 144 s^3 t + 144 s^2 t^2\nonumber\\
& + 70 s t^3 + 15 t^4) - 2 M^4 t (90 s^4 + 300 s^3 t + 353 s^2 t^2 + 191 s t^3 + 45 t^4)\nonumber\\
& + 4 M^6 (30 s^4 + 240 s^3 t + 400 s^2 t^2 + 261 s t^3 + 70 t^4)\Big],
\end{align}

\item{The bubble coefficient:}
\begin{align}
&\text{Bu}^{\gamma}=\frac{\kappa^4\bra{p_1}p_2|p_3]^2}{7680 \pi^2 (-4 M^2 + t)^2 (M^4 - 2 M^2 s + s (s + t))}\Big[452 M^8 + M^6 (-904 s + 944 t)\nonumber\\
& + M^4 (452 s^2 + 484 s t - 973 t^2) - t^2 (13 s^2 + 13 s t + 30 t^2)\nonumber\\
& + 2 M^2 t (-8 s^2 + 5 s t+ 150 t^2)\Big].
\end{align}

\end{itemize}
For the photon case, the one-loop coefficients for the $(+-)$ helicity configuration could be obtained simply by replacing  $\bra{p_1}p_2|p_3]^2$ with  $\bra{p_3}p_2|p_1]^2$. One potential confusion of the above results is that while the photon helicity configurations (e.g., the helicity configuration $(-+)$, $(+-)$, etc) are written in terms of the in-in formalism, which means that all momenta flow in and is much more common in the modern scattering amplitude community, we have adopted the standard \emph{in-out formalism} to present the explicit expressions of various one-loop coefficients, in which we have two particles incoming and two particles outgoing. This choice introduces extra minus sign in various places involving the spinor chain $\bra{p_1}p_2|p_3]^2$, when compared to expressions of the in-in formalism.\footnote{One way to see this extra minus sign goes back to the explicit realization of the dotted and undotted Weyl spinors. Following Ref.~\cite{Henn:2014yza}, we have 
\begin{align*}
\lambda^{\alpha}=\frac{1}{\sqrt{p^0+p^3}} \left(
\begin{array}{c}
p^0+p^3\\
p^1+ip^2
\end{array}
\right),\qquad\qquad
\tilde{\lambda}^{\dot{\alpha}}=\frac{1}{\sqrt{p^0+p^3}} \left(
\begin{array}{c}
p^0+p^3\\
p^1-ip^2
\end{array}
\right),
\end{align*}
with $p^0$, $p^1$, $p^2$ and $p^3$ the four components of the four-momentum $p^{\mu}$. When reverting the direction of the momentum flow by doing the replacement $p^{\mu} \to -p^{\mu}$, it is straightforward to see that 
\begin{align*}
\lambda^{\alpha}\to i \lambda^{\alpha}, \qquad\qquad \tilde{\lambda}^{\dot{\alpha}}\to i \tilde{\lambda}^{\dot{\alpha}}.
\end{align*}
The extra minus sign in various spinor-chains squared then follows straightforwardly.} 
This hybrid convention would be helpful when comparing our results with those in previous literature.

When compared with the results of Ref.~\cite{Bjerrum-Bohr:2014zsa,Bjerrum-Bohr:2016hpa}, one could find out that the inclusion of the massless-scalar double-cut diagrams and photon double-cut diagrams changes both the triangle coefficient $\text{T}^{\eta}$ and the bubble coefficient $\text{Bu}^{\eta}$. It is also interesting to note that the Bern-Carrasco-Johansson (BCJ) -inspired relation
\begin{equation}
\frac{\text{Bo}^{\eta}}{M^2-s}+\frac{\text{Bo}'^{\eta}}{M^2-u}=\text{T}^{\eta},\qquad \text{for } \eta=\varphi, \gamma,
\end{equation}
proposed in Ref.~\cite{Bjerrum-Bohr:2014zsa,Bjerrum-Bohr:2016hpa} no longer holds now. But we do have checked explicitly that the above relation still holds if only the graviton double-cut diagrams are taken into consideration.

\subsection{Low-Energy Limit}
\label{LEL}
In the previous part of Section \ref{OLA} we have provided all the relevant one-loop coefficients of gravitational scattering of the massless scalar and photon from a massive scalar. To figure out the long-range effects of scattering processes of quantum gravity, we need to first compute the low-energy limit of the scattering amplitude, in which we have $s\to M^2+2M\omega$ and $t\to -\textbf{q}^2$ for the kinematic invariants, and $\bra{p_1}p_2|p_3]^2 (\bra{p_3}p_2|p_1]^2) \to4M^2\omega^2$ for the spinor chains. The low-energy limits of various one-loop coefficients are given as follows:
\begin{align}
&\text{Bo}^{\varphi}=\text{Bo}'^{\varphi}=\text{Bo}^{\gamma}=\text{Bo}'^{\gamma}=\frac{\kappa^4}{16\pi^2}M^4\omega^4,\nonumber\\
&\text{T}^{\varphi}=\text{T}^{\gamma}=-\frac{\kappa^4}{16\pi^2}M^2\omega^2\textbf{q}^2,\qquad \text{T}'^{\varphi}=\text{T}'^{\gamma}=-\frac{15\kappa^4}{256\pi^2}M^4\omega^2,\nonumber\\
&\text{Bu}^{\varphi}=\frac{371\kappa^4}{7680\pi^2}M^2\omega^2 \qquad\text{Bu}^{\gamma}=\frac{113\kappa^4}{7680\pi^2}M^2\omega^2.
\end{align}
It is straightforward to show that the above results for the photon one-loop coefficients hold regardless of whether the photon helicity configuration is chosen to be $(-+)$ or $(+-)$.

On the other hand, the low-energy limits of the scalar integrals are found to be (see Appendix \ref{UI} for more details):
\begin{align}
&I_4(t,s)+I_4(t,u)=-i\frac{1}{\textbf{q}^2}\frac{2\pi}{M\omega}\log(\frac{\textbf{q}^2}{M^2}),\\
&I_3(t,0)=\frac{1}{2\textbf{q}^2}\log^2\left(\frac{\textbf{q}^2}{\mu^2}\right),\\
&I_3(t,M^2)=-\frac{1}{2M^2}\left(\log\left(\frac{\textbf{q}^2}{M^2}\right)+\frac{\pi^2M}{|\textbf{q}|}\right),\\
&I_2(t,0)=2-\log\left(\frac{\textbf{q}^2}{\mu^2}\right).
\end{align}
Here $\mu^2$ is the mass scale used in dimensional regularization. It is worthwhile to note that the dimensional regularization is adopted to handle not only ultraviolet divergences but also infrared divergences. Temporarily, we shall label $\mu^2$ resulting from ultraviolet divergences and infrared divergences as $\mu_{\text{UV}}^2$ and $\mu_{\text{IR}}^2$ respectively. Certainly, physical observables should not depend on $\mu_{\text{UV}}^2$, but they can depend on $\mu_{\text{IR}}^2$. For instance, in the calculation of inclusive cross sections including massless particle, $\mu_{\text{IR}}$ is the detector threshold, representing the resolution of the measurement. We have not done an explicit calculation by taking the detector resolution into consideration at this stage. It is pointed out by Ref.~\cite{Bjerrum-Bohr:2014zsa,Bjerrum-Bohr:2016hpa} that a full analysis of the impact of detector resolution is complicated. In the absence of such calculations, we simply replace all $\mu$'s in the logarithms by an infrared scale $1/r_0$, which is also the option taken by Ref.~\cite{Bjerrum-Bohr:2014zsa,Bjerrum-Bohr:2016hpa}. 

Provided with the above results, the total gravitational scattering amplitude
\begin{equation}
\mathcal{M}_{\eta}=\mathcal{M}^{(0)}_{\eta}+\mathcal{M}^{(1)}_{\eta},
\end{equation}
then has the low-energy expansion
\begin{align}
&\mathcal{M}_{\eta}=\kappa^2\frac{M^2\omega^2}{\textbf{q}^2}+\kappa^4\frac{15M^3\omega^2}{512|\textbf{q}|} +\kappa^4\frac{15M^2\omega^2}{512\pi^2}\log\left(\frac{\textbf{q}^2}{M^2}\right)-\kappa^4\frac{M^2\omega^2}{32\pi^2}\log^2\left(\frac{\textbf{q}^2}{\mu^2}\right)\nonumber\\
&-\kappa^4\frac{\text{bu}^{\eta}}{(8\pi)^2}M^2\omega^2\log\left(\frac{\textbf{q}^2}{\mu^2}\right)-\kappa^4\frac{M^3\omega^3}{8\pi}\frac{i}{\textbf{q}^2}\log\left(\frac{\textbf{q}^2}{M^2}\right),
\end{align}
where $\text{bu}^{\eta}$ is related to $\text{Bu}^{\eta}$ by
\begin{equation}
\text{bu}^{\eta}=\frac{(8\pi)^2}{\kappa^4M^2\omega^2}\text{Bu}^{\eta}.
\end{equation}
For the massless scalar $\varphi$ and the photon $\gamma$, $\text{bu}^{\eta}$s are given explicitly by 
\begin{equation}
\text{bu}^{\varphi}=\frac{371}{120},\qquad \text{bu}^{\gamma}=\frac{113}{120}.
\end{equation}
As mentioned before, the mass scale $\mu^2$ is introduced by dimensional regularization. In the following discussions of quantum corrections to the Newtonian potential in the coordinate space and bending of light, we shall simply replace the scale $\mu$ by the infrared scale $1/r_0$.

Before moving on, we would like to remark on the logarithm-squared part $\log^2\left(\textbf{q}^2/\mu^2\right)$. It comes from the massless triangle integral $I_3(t,0)$ and is related to infrared divergences. It is noteworthy that it is not canceled by the extra contributions from the massless-scalar and photon double-cut diagrams. In fact, the general structure of the one-loop amplitudes suggested in Ref.~\cite{Bjerrum-Bohr:2014zsa,Bjerrum-Bohr:2016hpa} remain intact even with the extra massless-scalar and photon double-cut contributions. None of them are canceled. 

\section{Bending of Light}
\label{BOL}
Now, we are ready to calculate quantum corrections to gravitational bending of light, which is perhaps the most famous experimental verification of Einstein's general relativity. To relate the microscopic scattering amplitude data to the macroscopic bending angle, one could use either the eikonal approximation or the semiclassical potential method. We recommend Ref.~\cite{Bjerrum-Bohr:2016hpa} for a recent discussion on this issue. In this article, we shall simply use the semiclassical potential method.

Let's first calculate the quantum Newtonian potential $V_\eta(r)$. The classical Newtonian potential $V_{\eta}^{(0)}(r)$ is given in Section \ref{TLA}, while the quantum corrections $V_{\eta}^{(1)}(r)$ could be induced easily from Section \ref{LEL}. The total quantum Newtonian potential $V(r)=V_{\eta}^{(0)}(r)+V_{\eta}^{(1)}(r)$ is then given by
\begin{align}
&\tilde{V}_{\eta}(\textbf{q})=-\frac{\kappa^2}{4}\frac{M\omega}{\textbf{q}^2}-\kappa^4\frac{15M^2\omega}{2048|\textbf{q}|} -\kappa^4\frac{15M\omega}{2048\pi^2}\log\left(\frac{\textbf{q}^2}{M^2}\right)+\kappa^4\frac{M\omega}{128\pi^2}\log^2\left(\frac{\textbf{q}^2}{\mu^2}\right)\nonumber\\
&+\kappa^4\frac{\text{bu}^{\eta}}{256\pi^2}M\omega\log\left(\frac{\textbf{q}^2}{\mu^2}\right),\\
&\!\!\!\!\!\!\!\!\!\!\!\!\!\Longrightarrow V_{\eta}(r)=\int\frac{\mathrm{d}^3\textbf{q}}{(2\pi)^3}\exp(i\textbf{q}\cdot\textbf{r})\tilde{V}_{\eta}(\textbf{q})\nonumber\\
&=-\frac{2GM\omega}{r}-\frac{15}{4}\frac{G^2M^2\omega}{r^2}+\frac{-8\text{bu}^{\eta}+15+64\log(r/r_0)}{4\pi}\frac{\hbar G^2M\omega}{r^3},
\end{align}
where $r_0$ is an undetermined infrared mass scale.

Provided with the above semiclassical potential, the bending angles of the massless scalar and the photon could be derived as follows
\begin{align}
\theta_{\eta}&=\frac{b}{\omega}\int_{-\infty}^{\infty}\mathrm{d}u\frac{V'_{\eta}(b\sqrt{1+u^2})}{\sqrt{1+u^2}}\nonumber\\
&=\frac{4GM}{b}+\frac{15}{4}\frac{G^2M^2\pi}{b^2}+\frac{8\text{bu}^{\eta}-47+64\log(2r_0/b)}{\pi}\frac{G^2\hbar M}{b^3}.
\end{align}
The above result is expressed in terms of the gauge-invariant impact parameter $b$. See Ref.~\cite{Bodenner:2003} for a nice discussion on the gauge-invariant parametrization of the gravitational bending angle. The first two terms give the correct classical bending angle, including the first post-Newtonian correction, while the last term gives the quantum gravity effect. The difference between bending angles of the photon and the massless scalar is then given by
\begin{equation}
\theta_{\gamma}-\theta_{\varphi}=\frac{8(\text{bu}^{\gamma}-\text{bu}^{\varphi})}{\pi}\frac{G^2\hbar M}{b^3},
\end{equation}
with the $\text{bu}^{\gamma}-\text{bu}^{\varphi}$ coefficient given by
\begin{equation}
\text{bu}^{\gamma}-\text{bu}^{\varphi}=-\frac{43}{20}.
\end{equation}
As a result, the quantum violation of some classical formulations of equivalence principle suggested in Ref.~\cite{Bjerrum-Bohr:2014zsa,Bjerrum-Bohr:2016hpa} remains to be valid after including the massless-scalar and photon double-cut contributions. 

Last but not least, it is noted in Ref.~\cite{Bjerrum-Bohr:2014zsa,Bjerrum-Bohr:2016hpa} that their calculations could reproduce the classical first post-Newtonian correction exactly. It is interesting to see how this comes into being even without including the massless-scalar and photon double-cut contributions. Given the general structure of the gravitational scattering amplitudes and the explicit expressions of the standard scalar integrals, it is straightforward to see that only the massive triangle integral $I_3(t,M^2)$ contributes to the post-Newtonian correction. As the massless-scalar and photon double-cut contributions do not contain any contribution of $I_3(t,M^2)$, they do not contribute to the post-Newtonian correction.

\section{Further Directions}
\label{FD}
In this article, we have derived the one-loop amplitudes for the massless scalar and the photon scattering gravitationally from a massive scalar object, and extracted the corresponding gravitational bending angles. We consider not only the graviton double-cut contributions, which have been studied in Ref.~\cite{Bjerrum-Bohr:2014zsa,Bjerrum-Bohr:2016hpa}, but also the extra contributions from the massless-scalar and photon double-cut diagrams. The final results for the gravitational bending angles are given by
\begin{align}
\theta_{\eta}=\frac{4GM}{b}+\frac{15}{4}\frac{G^2M^2\pi}{b^2}+\frac{8\text{bu}^{\eta}-47+64\log(2r_0/b)}{\pi}\frac{G^2\hbar M}{b^3},
\end{align}
with 
\begin{align}
\text{bu}^{\varphi}=\frac{371}{120}, \qquad\qquad \text{bu}^{\gamma}=\frac{113}{120},
\end{align}
for the massless scalar $\varphi$ and photon $\gamma$ respectively. In spite of the quantitative differences in various one-loop coefficients, quantum Newtonian potentials and gravitational bending angles, we find that the general structures of one-loop amplitudes suggested in Ref.~\cite{Bjerrum-Bohr:2014zsa,Bjerrum-Bohr:2016hpa} remain the same. So does the quantum discrepancy of some classical formulations of the equivalence principle.

There are several directions to extend our analysis:

1.~It is interesting to compute quantum corrections to other general relativity observables, such as the gravitational red shift, the Shapiro time delay, etc. Especially, it is shown recently that the Shapiro time delay puts extra constraints on the graviton three-point couplings \cite{Camanho:2014apa}, and the popular de Rham-Gabadadze-Tolley massive gravity \cite{Camanho:2016opx}. It is theoretically interesting to study its possible quantum corrections. 

2.~It is interesting to compute the one-loop gravitational amplitudes for massless spin-$1/2$, spin-$3/2$ and spin-$2$ particles. There has been some progress in this direction. For instance, the graviton double-cut contributions of the massless spin-$1/2$ particle have been derived in Ref.~\cite{Bjerrum-Bohr:2016hpa} using the on-shell unitarity method.  

3.~It is recognized that the first post-Newtonian correction appears in the one-loop calculations. A natural guess is that the post-Newtonian corrections of higher orders should correspond to the multi-loop amplitudes. The second post-Newtonian correction could be calculated using the method of Ref.~\cite{Bodenner:2003} from the general-relativity side. To validate the above conjecture, one has to then carry out a two-loop calculation from the quantum-field-theory side.

Hopefully, we will return to these issues in further publications.

\acknowledgments

DB would like to thank Prof.~Yue-Liang Wu for enlightening discussions and warm-hearted encouragement, and Prof.~N.E.J.~Bjerrum-Bohr for helpful communications. DB and YH would like to express their deep thanks to Prof.~N.E.J.~Bjerrum-Bohr, Prof.~John F.~Donoghue, Prof.~Barry R.~Holstein, Dr.~Ludovic Plant\'e, and Prof.~Pierre Vanhove for valuable comments and suggestions on the manuscript. Also, DB would like to thank Zhen Fang and Rui Zhang for helpful discussions during preparations for the revised manuscript. Last but not least, DB and YH would like to thank the anonymous referee for his/her careful and constructive review.

\appendix
\section{Notations and Conventions}
\label{AppendixA}
This appendix gives a comprehensive description of the notations and conventions used in this paper, and we hope that it could help the readers reproduce our results by themselves. The following notations and conventions could also be easily implemented in various Mathematica packages like xAct \cite{xAct}, FeynRules \cite{Alloul:2013bka}, FeynArts \cite{Hahn:2000kx}, FormCalc \cite{Hahn:2016ebn} and S@M \cite{Maitre:2007jq}, which (semi-)automatize the symbolic calculations. 

\begin{itemize}
\item Units:
\begin{equation}
\hbar=c=1.
\end{equation}

\item Metric signature: 
\begin{equation}
g_{\mu\nu}=\text{diag}(+,-,-,-).
\end{equation}
In other words, we adopt the mostly minus metric signature, which is commonly used in perturbative calculations of particle physics.

\item Levi-Civita tensor:
\begin{equation}
\epsilon^{\mu\nu\rho\sigma}=
\begin{cases}
1 & \quad\text{if $\{\mu,\nu,\rho,\sigma\}$ is an even permutation of $\{0,1,2,3\}$},\\
-1& \quad\text{if $\{\mu,\nu,\rho,\sigma\}$ is an odd permutation of $\{0,1,2,3\}$},\\
0 & \quad\text{others.}
\end{cases}
\end{equation}

\item Fourier transformations: In four dimensions, the Fourier transformations are defined as
\begin{align}
&f(x)=\int\!\frac{\mathrm{d}^4k}{(2\pi)^4}\exp(-ik\cdot x)\tilde{f}(k),\\
&\tilde{f}(k)=\int\!\mathrm{d}^4x\exp(ik\cdot x)f(x).
\end{align}
In three dimensions, the Fourier transformations are defined as
\begin{align}
&f(x)=\int\!\frac{\mathrm{d}^3\textbf{k}}{(2\pi)^3}\exp(i\textbf{k}\cdot \textbf{x})\tilde{f}(\textbf{k}),\\
&\tilde{f}(\textbf{k})=\int\!\mathrm{d}^3\textbf{k}\exp(-i\textbf{k}\cdot\textbf{x})f(\textbf{x}).
\end{align}
In this article, the following relations are useful in deriving the classical and quantum Newtonian potential in coordinate space
\begin{align}
&\int\!\frac{\mathrm{d}^3\textbf{q}}{(2\pi)^3}\exp(i\textbf{q}\cdot\textbf{r})\frac{1}{\textbf{q}^2}=\frac{1}{4\pi r},\label{3DFT1}\\
&\int\!\frac{\mathrm{d}^3\textbf{q}}{(2\pi)^3}\exp(i\textbf{q}\cdot\textbf{r})\frac{1}{|\textbf{q}|}=\frac{1}{2\pi^2r^2},\\
&\int\!\frac{\mathrm{d}^3\textbf{q}}{(2\pi)^3}\exp(i\textbf{q}\cdot\textbf{r})\log(\textbf{q}^2)=-\frac{1}{2\pi r^3},\\\
&\int\!\frac{\mathrm{d}^3\textbf{q}}{(2\pi)^3}\exp(i\textbf{q}\cdot\textbf{r})\log^2\left(\frac{\textbf{q}^2}{\mu^2}\right)=\frac{2\log\frac{r}{r_0}}{\pi r^3},\qquad \text{with $r_0=e^{1-\gamma_E}\mu^{-1}$}.
\label{3DFT4}
\end{align}
Here $\gamma_E$ is the Euler constant. When deriving the above three-dimensional Fourier transformation relations, it is convenient to first do the calculations in the general $D$ dimensions using, say, Mathematica, and then Taylor expand the obtained results around $d=3$. And this is how the Euler constant $\gamma_E$ comes into being in Eq.~\eqref{3DFT4}. 

\item Christoffel symbols:
\begin{equation}
\Gamma^{\mu}_{\alpha\beta}=\frac{1}{2}g^{\mu\nu}(\partial_{\alpha}g_{\nu\beta}+\partial_{\beta}g_{\nu\alpha}-\partial_{\nu}g_{\alpha\beta}).
\end{equation}

\item Riemann tensor: 
\begin{equation}
R_{\delta\gamma\beta}^{\ \ \ \ \alpha}=\partial_{\gamma}\Gamma^{\alpha}_{\beta\delta}-\partial_{\delta}\Gamma^{\alpha}_{\beta\gamma}+\Gamma^{\alpha}_{\mu\gamma}\Gamma^{\mu}_{\beta\delta}-\Gamma^{\alpha}_{\mu\delta}\Gamma^{\mu}_{\beta\gamma}.
\end{equation}
This is the convention adopted by Wald's textbook \cite{Wald:1984rg} and the Mathematica package xAct \cite{xAct}. 

\item Einstein-Hilbert action:
\begin{equation}
\mathcal{S}_{\text{EH}}=\int \mathrm{d}^4x \sqrt{-g}\left(-\frac{2}{\kappa^2}R\right),
\end{equation}
with $\kappa^2=32\pi G$, where $G$ is Newton's constant. The Einstein-Hilbert action here is different from that in Wald's textbook by a minus sign. This is due to the fact that we adopt the mostly minus metric signature, rather than the mostly plus metric signature employed by Wald. More explicitly, $R=g^{\mu\nu}R_{\mu\nu}=g^{\mu\nu}R_{\mu\alpha\nu}^{\ \ \ \ \alpha}$. Given the above definitions of Christoffel symbols and Riemann tensor, one only needs to replace $g_{\mu\nu}$ with $-g_{\mu\nu}$ in order to convert expressions in the mostly plus metric signature to those in the mostly minus metric signature. That gives the minus sign in our Einstein-Hilbert action.

\item S-matrix:
\begin{align}
&S=\mathbf{1}+i T,\nonumber\\
&\braket{\textbf{k}_1\textbf{k}_2\cdots|iT|\textbf{p}_1\textbf{p}_2\cdots}\equiv(2\pi)^4\delta^{(4)}(\sum p_i-\sum k_f)\cdot i \mathcal{M}(\textbf{p}_1\textbf{p}_2\cdots\to\textbf{k}_1\textbf{k}_2\cdots).\nonumber
\end{align}
Here the four-momentum is denoted by the normal letter, while the three-momentum is denoted by the boldface letter. For the nonrelativistic scattering in the external field of a massive object, the semi-classical potential function $V(\textbf{r})$ is given by
\begin{align}
\braket{\textbf{p}_f|iT|\textbf{p}_i}&\equiv(2\pi)^4\delta^{(4)}(p_i-p_f)\cdot i \mathcal{M}(q)\nonumber\\
&=-(2\pi)\delta(E_i-E_f)\cdot i \tilde{V}(\textbf{q}),
\label{DefinitionNewtonianPotential1}
\end{align}
with $p_i$ and $p_f$ being the incoming and outgoing four-momentum, and $q = p_f-p_i$. One could see further that 
\begin{align}
\tilde{V}(\textbf{q})=-\frac{1}{2M}\frac{1}{2\omega}\mathcal{M},
\label{DefinitionNewtonianPotential2}
\end{align}
where $\omega$ is the energy of the massless particle (e.g., the photon $\gamma$) and $M$ is the mass of the massive object (e.g., the massive scalar $\Phi$). As a result,
\begin{align}
V(r)=-\frac{1}{2M}\frac{1}{2\omega}\int\frac{\mathrm{d}^3\textbf{q}}{(2\pi)^3}\exp(i\textbf{q}\cdot\textbf{r})\mathcal{M}.
\label{DefinitionNewtonianPotential3}
\end{align} 

\item Spinor-helicity formalism: We follow the convention of Henn and Plefka's textbook \cite{Henn:2014yza}. 

\item Feynman propagators: In this article, we quantize the gravity using background-field method following 't Hooft and Veltman \cite{tHooft:1974toh}. In the background-field method, fields are expanded with respect to arbitrary classical background fields, with only the quantized fields treated as dynamical. The gauge invariances of the quantized fields are then broken by a choice of quantum gauge in a way that the resulting action is still invariant under {\it{background}} gauge transformations. Explicitly, we adopt the standard de Donder gauge and Feynman gauge to fix the quantum gauge invariances of graviton and photon respectively. For a detailed description of the quantization of the Einstein-Maxwell system, we recommend  Ref.~\cite{Deser:1974cz}. The relevant propagators in the Minkowski spacetime could then be found as follows:
\begin{align}
&\!\!\!\!\!\!\!\!\!\!\!\!\text{- Scalar propagator}\qquad\qquad\qquad\qquad \frac{i}{q^2-m^2+i\epsilon},\\
&\!\!\!\!\!\!\!\!\!\!\!\!\text{- Photon propagator}\qquad\qquad\qquad\qquad \frac{-i \eta_{\mu\nu}}{q^2+i\epsilon},\\
&\!\!\!\!\!\!\!\!\!\!\!\!\text{- Graviton propagator}\quad\quad \frac{1}{2}\Big[\eta_{\mu_1\mu_2}\eta_{\nu_1\nu_2}+\eta_{\mu_1\nu_2}\eta_{\nu_1\mu_2}-\eta_{\mu_1\nu_1}\eta_{\mu_2\nu_2}\Big]\frac{i}{q^2+i\epsilon},\\
&\!\!\!\!\!\!\!\!\!\!\!\!\text{- Graviton-ghost propagator}\qquad\qquad\qquad\qquad \frac{-i \eta_{\mu\nu}}{q^2+i\epsilon}.
\end{align}
We have attached along with this article a FeynArts model file containing all the above propagators as well as interacting vertices up to four points. It acts as the starting point of our automatic calculation routines for the one-loop scattering amplitudes.

\end{itemize}

\section{Useful Integrals}
\label{UI}
In this appendix, we summarize the explicit expressions of various scalar integrals that appear in our calculations. We adopt the notations of Ref.~\cite{Ellis:2007qk}:
\begin{align}
&I_2^D(p_1^2;m_1^2,m_2^2)=\frac{\mu^{4-D}}{i\pi^{\frac{D}{2}}r_{\Gamma}}\int\!\mathrm{d}^Dl\ \frac{1}{(l^2-m_1^2+i\epsilon)((l+q_1)^2-m_2^2+i\epsilon)},\\
&I_3^D(p_1^2,p_2^2,p_3^2;m_1^2,m_2^2,m_3^2)=\frac{\mu^{4-D}}{i\pi^{\frac{D}{2}}r_{\Gamma}}\nonumber\\
&\times\int\!\mathrm{d}^Dl\ \frac{1}{(l^2-m_1^2+i\epsilon)((l+q_1)^2-m_2^2+i\epsilon)((l+q_2)^2-m_3^2+i\epsilon)},\\
&I_4^D(p_1^2,p_2^2,p_3^2,p_4^2;s_{12},s_{23};m_1^2,m_2^2,m_3^2,m_4^2)=\frac{\mu^{4-D}}{i\pi^{\frac{D}{2}}r_{\Gamma}}\nonumber\\
&\times\int\!\mathrm{d}^Dl\ \frac{1}{(l^2-m_1^2+i\epsilon)((l+q_1)^2-m_2^2+i\epsilon)((l+q_2)^2-m_3^2+i\epsilon)((l+q_3)^2-m_4^4+i\epsilon)},
\end{align}
where $r_{\Gamma}=\Gamma(1-\epsilon)^2\Gamma(1+\epsilon)/\Gamma(1-2\epsilon)$, $q_n=\sum_{i=1}^np_i$, $q_0=0$, and $s_{ij}=(p_i+p_j)^2$.

The scalar integrals $I_4(t,s)$, $I_4(t,u)$, $I_3(t,0)$, $I_3(t,M^2)$ and $I_2(t,0)$ in this article are then given as follows
\begin{align}
&I_4(t,s)=I_4^D(t,0,s,M^2;0,M^2;0,0,0,M^2)\nonumber\\
&=-\frac{1}{t(M^2-s)}\left(\frac{\mu^2}{M^2}\right)^{\epsilon}\Bigg[\frac{2}{\epsilon^2}-\frac{1}{\epsilon}\left(2\log\frac{M^2-s}{M^2}+\log\frac{-t}{M^2}\right)\nonumber\\
&+2\log\frac{M^2-s}{M^2}\log\frac{-t}{M^2}-\frac{\pi^2}{2}+O(\epsilon)\Bigg],\label{I4ts}\\
&I_3(t,0)=I_3^D(t,0,0;0,0,0)=-\frac{1}{t\epsilon^2}-\frac{\log(-t/\mu^2)}{t\epsilon}-\frac{\log^2(-t/\mu^2)}{2t}+O(\epsilon),\\
&I_3(t,M^2)=I_3^D(t,M^2,M^2;0,0,M^2)\nonumber\\
&=\frac{1}{t\beta}\Bigg[\frac{2\pi^2}{3} + 2\text{Li}_2\left(\frac{\beta-1}{\beta+1}+\frac{1}{2}\log^2\left(\frac{\beta-1}{\beta+1}\right)\right)\Bigg],\qquad\text{with  $\beta^2=1-\frac{4M^2}{t}$,}\\
&I_2(t,0)=I_2^D(t,0,0)=\frac{1}{\epsilon}+2-\log\left(\frac{-t}{\mu^2}\right)+O(\epsilon).
\end{align}
The scalar integral $I_4(t,u)$ could be obtained by replacing $s$ with $u$ in Eq.~\eqref{I4ts}.



\begin{thebibliography}{99}

  %\cite{Bjerrum-Bohr:2014zsa}
\bibitem{Bjerrum-Bohr:2014zsa} 
  N.~E.~J.~Bjerrum-Bohr, J.~F.~Donoghue, B.~R.~Holstein, L.~Plant\'e and P.~Vanhove,
  ``Bending of Light in Quantum Gravity,''
  Phys.\ Rev.\ Lett.\  {\bf 114}, no. 6, 061301 (2015)
  %doi:10.1103/PhysRevLett.114.061301
  [arXiv:1410.7590 [hep-th]].
  %%CITATION = doi:10.1103/PhysRevLett.114.061301;%%
  %13 citations counted in INSPIRE as of 28 Nov 2016

%\cite{Bjerrum-Bohr:2016hpa}
\bibitem{Bjerrum-Bohr:2016hpa} 
  N.~E.~J.~Bjerrum-Bohr, J.~F.~Donoghue, B.~R.~Holstein, L.~Plant\'e and P.~Vanhove,
  ``Light-like Scattering in Quantum Gravity,''
  JHEP {\bf 1611}, 117 (2016)
  [arXiv:1609.07477 [hep-th]].
  %%CITATION = ARXIV:1609.07477;%%
  %7 citations counted in INSPIRE as of 28 Nov 2016

%\cite{Wu:2015wwa}
\bibitem{Wu:2015wwa} 
  Y.~L.~Wu,
  ``Quantum field theory of gravity with spin and scaling gauge invariance and spacetime dynamics with quantum inflation,''
  Phys.\ Rev.\ D {\bf 93}, no. 2, 024012 (2016)
  %doi:10.1103/PhysRevD.93.024012
  [arXiv:1506.01807 [hep-th]].
  %%CITATION = doi:10.1103/PhysRevD.93.024012;%%
  %3 citations counted in INSPIRE as of 28 Nov 2016
  
 %\cite{Wu:2015hoa}
\bibitem{Wu:2015hoa} 
  Y.~L.~Wu,
  ``Theory of Quantum Gravity Beyond Einstein and Space-time Dynamics with Quantum Inflation,''
  Int.\ J.\ Mod.\ Phys.\ A {\bf 30}, no. 28n29, 1545002 (2015)
  %doi:10.1142/S0217751X15450025
  [arXiv:1510.04720 [hep-th]].
  %%CITATION = doi:10.1142/S0217751X15450025;%%
  %1 citations counted in INSPIRE as of 28 Nov 2016

%\cite{Donoghue:1993eb}
\bibitem{Donoghue:1993eb} 
  J.~F.~Donoghue,
  ``Leading quantum correction to the Newtonian potential,''
  Phys.\ Rev.\ Lett.\  {\bf 72}, 2996 (1994)
  %doi:10.1103/PhysRevLett.72.2996
  [gr-qc/9310024].
  %%CITATION = doi:10.1103/PhysRevLett.72.2996;%%
  %298 citations counted in INSPIRE as of 28 Nov 2016

%\cite{Donoghue:1994dn}
\bibitem{Donoghue:1994dn} 
  J.~F.~Donoghue,
  ``General relativity as an effective field theory: The leading quantum corrections,''
  Phys.\ Rev.\ D {\bf 50}, 3874 (1994)
  %doi:10.1103/PhysRevD.50.3874
  [gr-qc/9405057].
  %%CITATION = doi:10.1103/PhysRevD.50.3874;%%
  %552 citations counted in INSPIRE as of 28 Nov 2016
  
  %\cite{Muzinich:1995uj}
\bibitem{Muzinich:1995uj} 
  I.~J.~Muzinich and S.~Vokos,
  ``Long range forces in quantum gravity,''
  Phys.\ Rev.\ D {\bf 52}, 3472 (1995)
  %doi:10.1103/PhysRevD.52.3472
  [hep-th/9501083].
  %%CITATION = doi:10.1103/PhysRevD.52.3472;%%
  %75 citations counted in INSPIRE as of 10 Feb 2017
  
 %\cite{Hamber:1995cq}
\bibitem{Hamber:1995cq} 
  H.~W.~Hamber and S.~Liu,
  ``On the quantum corrections to the Newtonian potential,''
  Phys.\ Lett.\ B {\bf 357}, 51 (1995)
  %doi:10.1016/0370-2693(95)00790-R
  [hep-th/9505182].
  %%CITATION = doi:10.1016/0370-2693(95)00790-R;%%
  %106 citations counted in INSPIRE as of 10 Feb 2017
  
  %\cite{Akhundov:1996jd}
\bibitem{Akhundov:1996jd} 
  A.~A.~Akhundov, S.~Bellucci and A.~Shiekh,
  ``Gravitational interaction to one loop in effective quantum gravity,''
  Phys.\ Lett.\ B {\bf 395}, 16 (1997)
  %doi:10.1016/S0370-2693(96)01694-2
  [gr-qc/9611018].
  %%CITATION = doi:10.1016/S0370-2693(96)01694-2;%%
  %63 citations counted in INSPIRE as of 10 Feb 2017
  
  %\cite{Kazakov:2000mu}
\bibitem{Kazakov:2000mu} 
  K.~A.~Kazakov,
  ``On the notion of potential in quantum gravity,''
  Phys.\ Rev.\ D {\bf 63}, 044004 (2001)
  %doi:10.1103/PhysRevD.63.044004
  [hep-th/0009220].
  %%CITATION = doi:10.1103/PhysRevD.63.044004;%%
  %10 citations counted in INSPIRE as of 10 Feb 2017
  
  %\cite{BjerrumBohr:2002sx}
\bibitem{BjerrumBohr:2002sx} 
  N.~E.~J.~Bjerrum-Bohr,
  ``Leading quantum gravitational corrections to scalar QED,''
  Phys.\ Rev.\ D {\bf 66}, 084023 (2002)
  %doi:10.1103/PhysRevD.66.084023
  [hep-th/0206236].
  %%CITATION = doi:10.1103/PhysRevD.66.084023;%%
  %43 citations counted in INSPIRE as of 10 Feb 2017

%\cite{BjerrumBohr:2002kt}
\bibitem{BjerrumBohr:2002kt} 
  N.~E.~J.~Bjerrum-Bohr, J.~F.~Donoghue and B.~R.~Holstein,
  ``Quantum gravitational corrections to the nonrelativistic scattering potential of two masses,''
  Phys.\ Rev.\ D {\bf 67}, 084033 (2003)
  Erratum: [Phys.\ Rev.\ D {\bf 71}, 069903 (2005)]
  %doi:10.1103/PhysRevD.71.069903, 10.1103/PhysRevD.67.084033
  [hep-th/0211072].
  %%CITATION = doi:10.1103/PhysRevD.71.069903, 10.1103/PhysRevD.67.084033;%%
  %166 citations counted in INSPIRE as of 28 Nov 2016
  
  %\cite{BjerrumBohr:2002ks}
\bibitem{BjerrumBohr:2002ks} 
  N.~E.~J.~Bjerrum-Bohr, J.~F.~Donoghue and B.~R.~Holstein,
  ``Quantum corrections to the Schwarzschild and Kerr metrics,''
  Phys.\ Rev.\ D {\bf 68}, 084005 (2003)
  Erratum: [Phys.\ Rev.\ D {\bf 71}, 069904 (2005)]
  %doi:10.1103/PhysRevD.68.084005, 10.1103/PhysRevD.71.069904
  [hep-th/0211071].
  %%CITATION = doi:10.1103/PhysRevD.68.084005, 10.1103/PhysRevD.71.069904;%%
  %101 citations counted in INSPIRE as of 10 Feb 2017

%\cite{Khriplovich:2002bt}
\bibitem{Khriplovich:2002bt} 
  I.~B.~Khriplovich and G.~G.~Kirilin,
  ``Quantum power correction to the Newton law,''
  J.\ Exp.\ Theor.\ Phys.\  {\bf 95}, no. 6, 981 (2002)
  [Zh.\ Eksp.\ Teor.\ Fiz.\  {\bf 122}, no. 6, 1139 (2002)]
  %doi:10.1134/1.1537290
  [gr-qc/0207118].
  %%CITATION = doi:10.1134/1.1537290;%%
  %62 citations counted in INSPIRE as of 10 Feb 2017
  
  %\cite{Khriplovich:2004cx}
\bibitem{Khriplovich:2004cx} 
  I.~B.~Khriplovich and G.~G.~Kirilin,
  ``Quantum long range interactions in general relativity,''
  J.\ Exp.\ Theor.\ Phys.\  {\bf 98}, 1063 (2004)
  [Zh.\ Eksp.\ Teor.\ Fiz.\  {\bf 125}, 1219 (2004)]
  %doi:10.1134/1.1777618
  [gr-qc/0402018].
  %%CITATION = doi:10.1134/1.1777618;%%
  %34 citations counted in INSPIRE as of 10 Feb 2017
  
  %\cite{Satz:2004hf}
\bibitem{Satz:2004hf} 
  A.~Satz, F.~D.~Mazzitelli and E.~Alvarez,
  ``Vacuum polarization around stars: Nonlocal approximation,''
  Phys.\ Rev.\ D {\bf 71}, 064001 (2005)
  %doi:10.1103/PhysRevD.71.064001
  [gr-qc/0411046].
  %%CITATION = doi:10.1103/PhysRevD.71.064001;%%
  %29 citations counted in INSPIRE as of 10 Feb 2017

  
  %\cite{Holstein:2008sx}
\bibitem{Holstein:2008sx} 
  B.~R.~Holstein and A.~Ross,
  ``Spin Effects in Long Range Gravitational Scattering,''
  arXiv:0802.0716 [hep-ph].
  %%CITATION = ARXIV:0802.0716;%%
  %18 citations counted in INSPIRE as of 28 Nov 2016
  
  %\cite{Park:2010pj}
\bibitem{Park:2010pj} 
  S.~Park and R.~P.~Woodard,
  ``Solving the Effective Field Equations for the Newtonian Potential,''
  Class.\ Quant.\ Grav.\  {\bf 27}, 245008 (2010)
  %doi:10.1088/0264-9381/27/24/245008
  [arXiv:1007.2662 [gr-qc]].
  %%CITATION = doi:10.1088/0264-9381/27/24/245008;%%
  %20 citations counted in INSPIRE as of 10 Feb 2017

%\cite{Marunovic:2011zw}
\bibitem{Marunovic:2011zw} 
  A.~Marunovic and T.~Prokopec,
  ``Time transients in the quantum corrected Newtonian potential induced by a massless nonminimally coupled scalar field,''
  Phys.\ Rev.\ D {\bf 83}, 104039 (2011)
  %doi:10.1103/PhysRevD.83.104039
  [arXiv:1101.5059 [gr-qc]].
  %%CITATION = doi:10.1103/PhysRevD.83.104039;%%
  %12 citations counted in INSPIRE as of 10 Feb 2017
  
  %\cite{Marunovic:2012pr}
\bibitem{Marunovic:2012pr} 
  A.~Marunovic and T.~Prokopec,
  ``Antiscreening in perturbative quantum gravity and resolving the Newtonian singularity,''
  Phys.\ Rev.\ D {\bf 87}, no. 10, 104027 (2013)
  %doi:10.1103/PhysRevD.87.104027
  [arXiv:1209.4779 [hep-th]].
  %%CITATION = doi:10.1103/PhysRevD.87.104027;%%
  %6 citations counted in INSPIRE as of 10 Feb 2017

%\cite{Burns:2014bva}
\bibitem{Burns:2014bva} 
  D.~Burns and A.~Pilaftsis,
  ``Matter Quantum Corrections to the Graviton Self-Energy and the Newtonian Potential,''
  Phys.\ Rev.\ D {\bf 91}, no. 6, 064047 (2015)
  %doi:10.1103/PhysRevD.91.064047
  [arXiv:1412.6021 [hep-th]].
  %%CITATION = doi:10.1103/PhysRevD.91.064047;%%
  %3 citations counted in INSPIRE as of 10 Feb 2017

%\cite{Frob:2016xte}
\bibitem{Frob:2016xte} 
  M.~B.~Fr0‹2b,
  ``Quantum gravitational corrections for spinning particles,''
  JHEP {\bf 1610}, 051 (2016)
  Erratum: [JHEP {\bf 1611}, 176 (2016)]
  %doi:10.1007/JHEP10(2016)051, 10.1007/JHEP11(2016)176
  [arXiv:1607.03129 [hep-th]].
  %%CITATION = doi:10.1007/JHEP10(2016)051, 10.1007/JHEP11(2016)176;%%
  %2 citations counted in INSPIRE as of 10 Feb 2017
  
  %\cite{Donoghue:2017pgk}
\bibitem{Donoghue:2017pgk} 
  J.~F.~Donoghue, M.~M.~Ivanov and A.~Shkerin,
  ``EPFL Lectures on General Relativity as a Quantum Field Theory,''
  arXiv:1702.00319 [hep-th].
  %%CITATION = ARXIV:1702.00319;%%

%\cite{Bjerrum-Bohr:2013bxa}
\bibitem{Bjerrum-Bohr:2013bxa} 
  N.~E.~J.~Bjerrum-Bohr, J.~F.~Donoghue and P.~Vanhove,
  ``On-shell Techniques and Universal Results in Quantum Gravity,''
  JHEP {\bf 1402}, 111 (2014)
  %doi:10.1007/JHEP02(2014)111
  [arXiv:1309.0804 [hep-th]].
  %%CITATION = doi:10.1007/JHEP02(2014)111;%%
  %17 citations counted in INSPIRE as of 28 Nov 2016
    
  %\cite{Holstein:2016cyq}
\bibitem{Holstein:2016cyq} 
  B.~R.~Holstein,
  ``Analytical On-shell Calculation of Low Energy Higher Order Scattering,''
  arXiv:1609.00714 [hep-ph].
  %%CITATION = ARXIV:1609.00714;%%
  %1 citations counted in INSPIRE as of 02 Dec 2016

  %\cite{Holstein:2016fxh}
\bibitem{Holstein:2016fxh} 
  B.~R.~Holstein,
  ``Analytical On-shell Calculation of Higher Order Scattering: Massive Particles,''
  arXiv:1610.07957 [hep-ph].
  %%CITATION = ARXIV:1610.07957;%%

%\cite{Holstein:2016cfx}
\bibitem{Holstein:2016cfx} 
  B.~R.~Holstein,
  ``Analytical On-shell Calculation of Higher Order Scattering: Massless Particles,''
  arXiv:1611.03074 [hep-ph].
  %%CITATION = ARXIV:1611.03074;%%
    
  \bibitem{xAct}
xAct: Efficient tensor computer algebra for the Wolfram Language, {\tt http://www.xact.es}.
  
  %\cite{Alloul:2013bka}
\bibitem{Alloul:2013bka} 
  A.~Alloul, N.~D.~Christensen, C.~Degrande, C.~Duhr and B.~Fuks,
  ``FeynRules  2.0 - A complete toolbox for tree-level phenomenology,''
  Comput.\ Phys.\ Commun.\  {\bf 185}, 2250 (2014)
  %doi:10.1016/j.cpc.2014.04.012
  [arXiv:1310.1921 [hep-ph]].
  %%CITATION = doi:10.1016/j.cpc.2014.04.012;%%
  %510 citations counted in INSPIRE as of 02 Dec 2016
  
  %\cite{Hahn:2000kx}
\bibitem{Hahn:2000kx} 
  T.~Hahn,
  ``Generating Feynman diagrams and amplitudes with FeynArts 3,''
  Comput.\ Phys.\ Commun.\  {\bf 140}, 418 (2001)
  %doi:10.1016/S0010-4655(01)00290-9
  [hep-ph/0012260].
  %%CITATION = doi:10.1016/S0010-4655(01)00290-9;%%
  %1129 citations counted in INSPIRE as of 02 Dec 2016

%\cite{Hahn:2016ebn}
\bibitem{Hahn:2016ebn} 
  T.~Hahn, S.~Paßehr and C.~Schappacher,
  ``FormCalc 9 and Extensions,''
  PoS LL {\bf 2016}, 068 (2016)
  %doi:10.1088/1742-6596/762/1/012065
  [arXiv:1604.04611 [hep-ph]].
  %%CITATION = doi:10.1088/1742-6596/762/1/012065;%%
  
  %\cite{Maitre:2007jq}
\bibitem{Maitre:2007jq} 
  D.~Maitre and P.~Mastrolia,
  ``S@M, a Mathematica Implementation of the Spinor-Helicity Formalism,''
  Comput.\ Phys.\ Commun.\  {\bf 179}, 501 (2008)
  %doi:10.1016/j.cpc.2008.05.002
  [arXiv:0710.5559 [hep-ph]].
  %%CITATION = doi:10.1016/j.cpc.2008.05.002;%%
  %74 citations counted in INSPIRE as of 02 Dec 2016
  
  %\cite{Passarino:1978jh}
\bibitem{Passarino:1978jh} 
  G.~Passarino and M.~J.~G.~Veltman,
  ``One Loop Corrections for e+ e- Annihilation Into mu+ mu- in the Weinberg Model,''
  Nucl.\ Phys.\ B {\bf 160}, 151 (1979).
  %doi:10.1016/0550-3213(79)90234-7
  %%CITATION = doi:10.1016/0550-3213(79)90234-7;%%
  %2080 citations counted in INSPIRE as of 13 Feb 2017


\bibitem{Bodenner:2003}
J.~Bodenner and C.M.~Will,
`` Deflection of light to second order: A tool for illustrating principles of general relativity,"
Am.~J.~Phys. {\bf 71}, 770 (2003).

%\cite{Camanho:2014apa}
\bibitem{Camanho:2014apa} 
  X.~O.~Camanho, J.~D.~Edelstein, J.~Maldacena and A.~Zhiboedov,
  ``Causality Constraints on Corrections to the Graviton Three-Point Coupling,''
  JHEP {\bf 1602}, 020 (2016)
  %doi:10.1007/JHEP02(2016)020
  [arXiv:1407.5597 [hep-th]].
  %%CITATION = doi:10.1007/JHEP02(2016)020;%%
  %126 citations counted in INSPIRE as of 02 Dec 2016

%\cite{Camanho:2016opx}
\bibitem{Camanho:2016opx} 
  X.~O.~Camanho, G.~L.~Gomez and R.~Rahman,
  ``Causality Constraints on Massive Gravity,''
  arXiv:1610.02033 [hep-th].
  %%CITATION = ARXIV:1610.02033;%%

%\cite{Wald:1984rg}
\bibitem{Wald:1984rg} 
  R.~M.~Wald,
  ``General Relativity,''
  Chicago, Usa: Univ. Pr. (1984) 491p.
  %doi:10.7208/chicago/9780226870373.001.0001
  %%CITATION = doi:10.7208/chicago/9780226870373.001.0001;%%
  %170 citations counted in INSPIRE as of 03 Dec 2016

%\cite{Henn:2014yza}
\bibitem{Henn:2014yza} 
  J.~M.~Henn and J.~C.~Plefka,
  ``Scattering Amplitudes in Gauge Theories,''
  Lect.\ Notes Phys.\  {\bf 883}, 1 (2014).
  %doi:978-3-642-54021-9, 10.1007/978-3-642-54022-6
  %%CITATION = doi:978-3-642-54021-9, 10.1007/978-3-642-54022-6;%%
  %34 citations counted in INSPIRE as of 15 Nov 2016

%\cite{tHooft:1974toh}
\bibitem{tHooft:1974toh} 
  G.~'t Hooft and M.~J.~G.~Veltman,
  ``One loop divergencies in the theory of gravitation,''
  Ann.\ Inst.\ H.\ Poincare Phys.\ Theor.\ A {\bf 20}, 69 (1974).
  %889 citations counted in INSPIRE as of 18 Feb 2017

%\cite{Deser:1974cz}
\bibitem{Deser:1974cz} 
  S.~Deser and P.~van Nieuwenhuizen,
  ``One Loop Divergences of Quantized Einstein-Maxwell Fields,''
  Phys.\ Rev.\ D {\bf 10}, 401 (1974).
  %doi:10.1103/PhysRevD.10.401
  %%CITATION = doi:10.1103/PhysRevD.10.401;%%
  %400 citations counted in INSPIRE as of 15 Feb 2017


%\cite{Ellis:2007qk}
\bibitem{Ellis:2007qk} 
  R.~K.~Ellis and G.~Zanderighi,
  ``Scalar one-loop integrals for QCD,''
  JHEP {\bf 0802}, 002 (2008)
  %doi:10.1088/1126-6708/2008/02/002
  [arXiv:0712.1851 [hep-ph]].
  %%CITATION = doi:10.1088/1126-6708/2008/02/002;%%
  %300 citations counted in INSPIRE as of 02 Dec 2016


% Please avoid comments such as "For a review'', "For some examples",
% "and references therein" or move them in the text. In general,
% please leave only references in the bibliography and move all
% accessory text in footnotes.

% Also, please have only one work for each \bibitem.


\end{thebibliography}
\end{document}